\documentclass[conference, a4paper]{IEEEtran}
\IEEEoverridecommandlockouts
\usepackage{cite}
\usepackage{amsmath,amssymb,amsfonts}
\DeclareMathOperator*{\argmax}{argmax} 
\usepackage{algorithmic}
\usepackage{graphicx}
\usepackage{textcomp}
\usepackage{xcolor}
\usepackage{multirow}
\usepackage{booktabs}
\usepackage{bbold}
\usepackage{url}
\usepackage{siunitx}
\usepackage{subcaption}
\usepackage{cuted}
\usepackage[caption=false]{subfig}
\usepackage[nolist]{acronym}
\usepackage[inline]{enumitem}
\usepackage{optidef}
\renewcommand{\baselinestretch}{0.954}
\usepackage[margin=0.72in]{geometry}       
\newtheorem{theorem}{Theorem}

\newtheorem{remark}{Remark}

\begin{document}

\makeatletter

\title{Trajectory-Independent Flexibility Envelopes of Energy-Constrained Systems with State-Dependent Losses
\thanks{This work was supported as a part of NCCR Automation, a National Centre of Competence in Research, funded by the Swiss National Science Foundation (grant number 51NF40\_225155) and by the Swiss Federal Office of Energy’s: “P+D” office and “SWEET” program.}
}

\author{Julie Rousseau, \IEEEmembership{Student Member IEEE}, 
Carlo Tajoli, \IEEEmembership{Student Member IEEE}, 
Hanmin Cai, \IEEEmembership{Member IEEE},\\
Philipp Heer, \IEEEmembership{Member IEEE}, 
Kristina Orehounig, 
Gabriela Hug, \IEEEmembership{Senior Member IEEE}

\thanks{C. Tajoli (ctajoli@ethz.ch) and G. Hug are with the Power Systems Laboratory at ETH Z\"{u}rich, Z\"{u}rich, Switzerland. P. Heer is with the Urban Energy Systems Laboratory, Empa, D\"{u}bendorf, Switzerland. K. Orehounig is with the Research Unit for Building Physics and Building Ecology, Vienna University of Technology, Vienna, Austria. J. Rousseau (jrousseau@ethz.ch) is both with the Power Systems Laboratory at ETH Z\"{u}rich, Z\"{u}rich, Switzerland, and the Urban Energy Systems Laboratory, Empa, D\"{u}bendorf, Switzerland.}

}

\maketitle

\begin{abstract}
    As non-dispatchable renewable power units become prominent in electric power grids, demand-side flexibility appears as a key element of future power systems' operation. Power and energy bounds are intuitive metrics to describe the flexibility of energy-constrained loads. However, to be used in operation, any power consumption trajectory fulfilling the power and energy bounds must necessarily fulfill the load's constraints. In this paper, we demonstrate that energy bounds defined as the minimum and maximum energy consumption potential of a load with state-dependent losses are Trajectory-Dependent (TD), i.e., for any energy value in the bounds a feasible power trajectory exists, but not all power trajectories enclosed in the energy envelopes satisfy the load's constraints. To guarantee the satisfaction of load constraints for all trajectories, we define Trajectory-Independent (TI) energy bounds. We present TI envelope formulations for individual loads, as well as physically coupled loads and assess the proposed formulations in a building heating system, a system with state-dependent losses. We find that using a TD envelope as energy bounds in operation may yield room temperature up to 3.8\textdegree C higher and 3.4\textdegree C lower than admissible. Overall, poorly insulated buildings observe a TI energy envelope that differs significantly from their TD envelope. 
\end{abstract}


\begin{acronym}[ML] 
    \acro{DTU}{Technical University of Denmark}
    \acro{MFPH}{Maximum Flexibility Provision Horizon}
\end{acronym}

\section{Introduction}

Power systems currently undergo significant transformations. Large fossil-fuel power plants that flexibly produce power to ensure the balance between power production and consumption are being decommissioned. At the same time, non-dispatchable renewable power units are being installed \cite{ReportIRENA2019}, but they cannot fully adapt their production to consumption. Hence, policymakers encourage power consumers to become more flexible, i.e., adapt a share of their consumption to the production of renewable units \cite{EUFlexibility}. 

However, integrating flexible consumers in the planning and operation of power systems is challenging, partly due to the diversity of flexible loads. Therefore, a unified flexibility representation, i.e., a way to represent various flexible loads in a standardized manner, is valuable \cite{LI2021100054}. 

Different unified flexibility representations that integrate loads' power and/or energy constraints have been proposed in the literature. The instantaneous power flexibility of a load describes its immediate power adaptation capability but does not describe its energy constraints over a longer horizon \cite{FINCK2019}. To account for the limited energy capacity of loads, their flexibility is described in \cite{riaz2022} using constant power values that can be sustained over a fixed duration, which neglects time-varying power profiles. Alternatively, by modeling loads as virtual batteries with losses, a load's energy and power constraints can be represented \cite{ULBIG2015}. However, in \cite{ULBIG2015} only state-independent losses are considered, excluding energy and power constraints of loads such as building heating systems. A virtual battery with state-dependent losses is presented in \cite{Hao2015}. Even though virtual batteries offer a modular representation of flexible loads, grid operators must provide a general representation accommodating all possible loads. 
Besides, flexible loads must share all technical private parameters with external stakeholders. 
Another unified flexibility representation is the concept of power and energy flexibility envelopes, which describe time-varying power and energy boundaries on the power consumption of a flexible resource \cite{Cai2021, Rousseau2023, LU202128, DHULST201579}. In this paradigm, each flexible load must quantify its flexibility envelopes, offering an intuitive metric for stakeholders. In this paper, we use this concept of power and energy flexibility envelopes to describe the flexibility of loads. 

Energy flexibility envelopes describe the feasible energy consumption of a load with two energy bounds: an upper bound, describing the maximum energy that the load can consume over a fixed horizon, e.g., a day, and a lower bound, representing its minimum energy consumption. Energy envelopes are intuitive to use in operation as they describe time-varying energy bounds. Nevertheless, they are rarely used in the literature in such an operational context. Indeed, there is no certainty that a load’s energy constraints are
satisfied if its power consumption trajectory lies within the power and energy flexibility bounds. More precisely, the energy envelope describes the set of feasible energy points, meaning that at least one feasible power trajectory leads to any energy consumption value, but it does not guarantee that all power trajectories satisfy the power and energy constraints of the load. Indeed, while energy flexibility envelopes are considered to be an intuitive metric, the authors in \cite{REYNDERS2018372} advocate caution when employing such envelopes as operational tools. 

Yet, existing studies have assumed the existence of such energy envelopes to build further methodologies, e.g., \cite{WEN2024110628, Mueller2019}.
By giving a counter-example, this paper demonstrates that energy flexibility envelopes, as defined in the literature, cannot be used in operation. 
Furthermore, we define a Trajectory-Independent (TI) energy flexibility envelope, which guarantees that all power trajectories satisfying power and energy bounds also satisfy the load's constraints. Such energy bounds can then be integrated into the operational tools of aggregators and grid operators.

In the remainder,
we present the notation, the system under study, and we mathematically state the goal of this paper in Section~\ref{sec:Prelim}. Then, we demonstrate that energy flexibility envelopes, presented in the literature and described in Section~\ref{sec:MaxPotential}, are Trajectory-Dependent (TD) for systems with state-dependent losses. Therefore, we propose a TI energy envelope for a single flexible load, described as a uni-dimensional system, in Section~\ref{sec:operationalPotential1D} and for a set of physically coupled flexible loads, described as a multi-dimensional system, in Section~\ref{sec:operationalPotentialMultiD}. In Section~\ref{sec:caseStudy}, we introduce the model of a heating system of a building, used in Section~\ref{sec:caseStudy} as a test case to evaluate the different formulations. In particular, we investigate how the formulations perform on different buildings. Finally, Section~\ref{sec:conclusion} concludes the paper and discusses future works. 
\section{Preliminaries}\label{sec:Prelim}
\subsection{Notations}

In the remainder of this paper, bold letters designate vectors or matrices. The notation $\boldsymbol{z}(t)$ indicates the value of the vector $\boldsymbol{z} \in \mathbb{R}^{N_z}$ at time instant $t$, while $\boldsymbol{z}_k$ designates the discretized value of the vector $\boldsymbol{z}$ at timestep $k$. The set $\mathcal{T}$ describes the continuous time horizon, while $\mathcal{I}$ and $\mathcal{T}_d$ designate the set of loads and the set of discretized timesteps, respectively. Integrals applied to vectors describe a component-wise integration. The capital letter $\text{E}$ consistently describes energy to distinguish from exponential terms, while the lowercase letter $p$ designates power. 

\subsection{System Under Study}
\label{subsec:System}

We restrict our work to lossy systems governed by linear state equations of the form: 
\begin{equation}
    \dfrac{d \boldsymbol{x} (t)}{dt} = \mathbf{A} \boldsymbol{x} (t) + \mathbf{B}_p \boldsymbol{p} (t) + \mathbf{B}_d \boldsymbol{d} (t),
    \label{eq:subID}
\end{equation}
where $\boldsymbol{x} \in \mathbb{R}^{N_x}$ designates the system's state, $\boldsymbol{p} \in \mathbb{R}^{N_p}_+$, the power inputs to the system, and $\boldsymbol{d} \in \mathbb{R}^{N_d}$, the additional inputs to the system which are assumed to be uncontrollable and independent of the states and power inputs. We assume that all power inputs are non-negative\footnote{The proofs presented in this paper can be extended to accommodate the case of non-positive power inputs, under the condition that non-positive power inputs lead to a decrease in state values, i.e., $\mathbf{B}_p \geq \boldsymbol{0}$.} and that all power inputs increase system states, i.e., $\mathbf{B}_p \geq \boldsymbol{0}$. 

We focus on systems containing non-positive diagonal elements in their matrix $\mathbf{A}$. Such systems suffer from state-dependent losses. Additionally, we limit our study to the case where all the off-diagonal components of $\mathbf{A}$ are non-negative\footnote{If all the elements of $\mathbf{A}$ are non-negative, the states grow exponentially. In this case, the integral form of Gronwall's lemma proves that the energy flexibility potential, presented in Section~\ref{subsec:MaxEnvelopeDef}, can be used in operation.}. This condition may also be referred to as the system being positive linear in the literature, or $\mathbf{A}$ being a Metzler matrix \cite{book_positive_systems}. Directed Laplacian matrices of networked systems verify this property \cite{blanes2022positivity}. The matrices describing buildings' thermal dynamics, which is the case study of this paper, follow these properties.

\subsection{Goal of the Paper}

The goal of this paper is to define a TI energy envelope, delimited by $\mathbf{E}_{\text{up}}$ and $\mathbf{E}_{\text{down}}$, such that if a power consumption trajectory fulfills the system's power constraints and lies in the energy envelopes, then the system's state constraints are satisfied. Mathematically speaking, if it holds that: 
\begin{equation}
    \mathbf{E}_{\text{down}} (t) \leq \int_{0}^t \boldsymbol{p} (\tau)  \text{d} \tau \leq \mathbf{E}_{\text{up}} (t), \quad \forall t \in \mathcal{T},
\end{equation}
and $\boldsymbol{p}$ fulfills the power constraints, it implies that: 
\begin{equation}
    \boldsymbol{x}_{\text{min}} \leq \boldsymbol{x} (t) \leq \boldsymbol{x}_{\text{max}}, \quad \forall t \in \mathcal{T}.
\end{equation}
We highlight that TI energy envelopes only aim to ensure state constraint satisfaction. 
Additional power constraints, such as ramping limits, are not considered in the model and must be imposed additionally.

\section{Trajectory-Dependent (TD) Energy Flexibility Potential}
\label{sec:MaxPotential}

The energy envelopes introduced in \cite{Cai2021, Rousseau2023, LU202128, DHULST201579} describe the energy flexibility potential of a load, i.e., how much energy a load can consume over a given horizon. We first rigorously define a load's energy flexibility potential and then explain the shortcomings of this metric. 

\subsection{Definition}
\label{subsec:MaxEnvelopeDef}
The energy flexibility potential of a flexible load is an energy consumption region delimited by two bounds, representing the minimum and maximum accumulated power over time that a load can consume over a certain time period. The upper energy bound $\mathbf{E}_{\text{up}}^{\text{\tiny TD}}$ corresponds to the optimal objective function's value of the problem: 
\begin{subequations}
\begin{align}
    \mathbf{E}_{\text{up}}^{\text{\tiny TD}} (t) = \max_{\boldsymbol{p}}  \quad & \int_{t \in \mathcal{T}} \boldsymbol{1}^\intercal \cdot \boldsymbol{p} (t)  \text{d} t, \\
    \text{s.t.} \quad & \text{(\ref{eq:subID})}, \label{eq:upperBound_1}\\
    & \boldsymbol{p}_{\text{min}} \leq \boldsymbol{p}(t) \leq \boldsymbol{p}_{\text{max}},  \quad \forall t \in \mathcal{T}, \label{eq:upperBound_2}\\
    & \boldsymbol{x}_{\text{min}} \leq \boldsymbol{x}(t) \leq \boldsymbol{x}_{\text{max}}, \quad \forall t \in \mathcal{T}.  \label{eq:upperBound_3}
\end{align}  
\label{eq:upperBound}
\end{subequations}
\hspace{-0.1cm}The lower energy bound $\mathbf{E}_{\text{down}}^{\text{\tiny TD}}$ can be computed similarly by instead minimizing the energy consumption. 

If $\mathbf{A}=0$, the system does not suffer from state-dependent losses, i.e., losses are independent of the system's state. Then, it can be proven that the load's energy flexibility potential is TI. However, this is not the case for systems affected by state-dependent losses, which is the focus of this paper.

\subsection{Limits in Operation}
\label{subsec:counterExampleMaxEnvelope}

\begin{figure}
    \centering
    \includegraphics[width = \columnwidth]{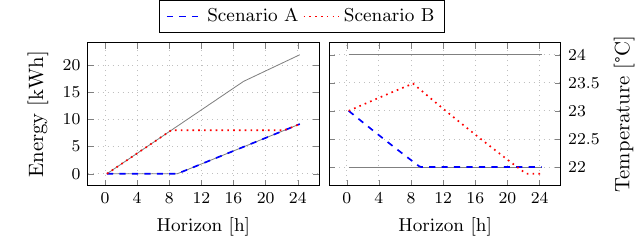}
    \caption{On the left, an example of a TD energy flexibility envelope of a one-zone building, defined by the gray lines, with two power consumption trajectories (Scenario A and B). On the right, their resulting room temperatures, with the temperature boundaries in gray.}
    \label{fig:envelopeOneZone}
\end{figure}

As an illustrative example, we consider the evolution of the room temperature in a building modeled as one thermal zone, which can be described by: 
\begin{equation}
    C \dfrac{d T }{dt} = \frac{1}{R} \left( T_a - T \right) (t) + p_{\text{th}} (t)  + d (t),
    \label{eq:oneZoneExmaple}
\end{equation}
where $C$ represents the building's heat capacity, $R$ is the building's thermal resistance, $T_a$ designates the ambient temperature, $p_{\text{th}}$ describes the heating thermal power, and $d$ denotes additional heat gains, e.g., from solar radiation or the building's inhabitants. The parameter values are specified in Table~\ref{tab:exampleOneZone} and correspond to the \textit{SwissHouse} archetype presented in \cite{energym}. In this example, the room temperature $T$ is the system's state. Fig.~\ref{fig:envelopeOneZone} depicts the energy flexibility potential of the system represented by the black lines in the plot on the left and computed based on (\ref{eq:upperBound}). 

\begin{table}[t]
    \centering
    \footnotesize
    \renewcommand{\arraystretch}{1.3}
    \setlength{\tabcolsep}{3pt}
    \caption{Thermal parameters of a one-zone building \cite{energym}, and the starting condition and boundaries for the heating operation.}
    \label{tab:exampleOneZone}

    \begin{minipage}{0.45\columnwidth}
        \centering
        \begin{tabular}{c c c}
            \toprule
            \textbf{Parameter} & Value & Unit \\ \midrule
            C & $20$ & $\text{MJ} / \text{K}$\\ 
            $1/R$ & 50  & $\text{W} / \text{K}$\\
            $d(t)$ & 0 & W \\
            $T_a (t)$ & 10 & °C \\
            \bottomrule
        \end{tabular}
    \end{minipage}%
    \begin{minipage}{0.45\columnwidth}
        \centering
        \begin{tabular}{c c c}
            \toprule
            \textbf{Parameter} & Value & Unit \\ \midrule
            $T_0$ & 23 & °C \\
            $\left[ T_{\text{min}}, T_{\text{max}}\right]$ & $\left[ 22, 24 \right]$ & °C \\
            $\left[ p_{\text{min}}, p_{\text{max}}\right]$ & $\left[ 0, 1 \right]$ & kW \\
            \bottomrule
        \end{tabular}
    \end{minipage}
\end{table}

To assess the validity of using energy flexibility potential envelopes in operation, we select two energy trajectories enclosed in the envelope, as represented in the left graph of Fig.~\ref{fig:envelopeOneZone}. Both trajectories also fulfill the power constraints of the device. In Scenario~A, the system consumes as little power and as late as possible. In Scenario~B, the system consumes as much power and as early as possible, leading to the same final energy consumption as in Scenario~A. Additionally, the right graph in Fig.~\ref{fig:envelopeOneZone} represents the resulting room temperature for both power consumption trajectories. While both trajectories fulfill the power and energy constraints, the room temperature in Scenario~B falls below the minimum acceptable temperature. This constitutes a counter-example showing that energy flexibility potential envelopes are TD. Using them in operation does not guarantee the satisfaction of the system's state constraints.

Analyzing losses explains the differences among the scenarios. Even though the final energy consumption of Scenario~A and~B are equal, the power trajectories generate different thermal losses to the ambient environment. The room temperature in Scenario~A is low throughout the horizon and only rises towards the end. This yields little thermal losses. In comparison, the room temperature in Scenario~B rises early, leading to higher thermal losses to the ambient environment throughout the horizon. Experiencing higher thermal losses while consuming as much final energy as in Scenario~A leads to less thermal energy stored in the building in Scenario~B, resulting in a lower final room temperature. Mathematically speaking, integrating (\ref{eq:oneZoneExmaple}) indicates that larger thermal losses yield a lower final temperature, assuming the same final energy consumption and initial room temperature. As a result, the room temperature in Scenario~B is smaller than the one in Scenario~A and falls below the minimum acceptable value. 

This constitutes a counter-example showing that when a system suffers from state-dependent losses, the energy flexibility envelope as defined in Section~\ref{subsec:MaxEnvelopeDef} is TD, i.e., that not all trajectories in the envelope guarantee the satisfaction of state constraints. Importantly, the scenarios considered in this example and displayed in Fig.~\ref{fig:envelopeOneZone} are reasonable scenarios in demand-side management, namely corresponding to the curtailment of loads at critical hours. In Scenario~A, critical hours would occur at the beginning of the horizon, while they would happen towards the end of the horizon in Scenario~B. Hence, it is critical to define TI energy envelopes for operational purposes.

\section{Trajectory-Independent (TI) Energy Flexibility: Uni-Dimensional Systems}
\label{sec:operationalPotential1D}

This section defines a TI envelope for uni-dimensional systems, i.e., systems with one state. We first provide the intuition used to build the TI envelopes, then derive a TI energy flexibility envelope formulation, and finally, we explain how to compute TI energy bounds. 

\subsection{Intuition}
 
As the example in Section~\ref{subsec:counterExampleMaxEnvelope} shows, different power trajectories may lead to similar final energy consumption values but different final states. 
In fact, the analytical solution of (\ref{eq:subID}) in the uni-dimensional case is: 
    \begin{equation}
        x(t) = e^{At} x(0) + \int_0^t e^{A (t - \tau) } \left( B_d d(\tau) + B_p p(\tau) \right) \text{d}\tau,
    \label{eq:analyticalSolution1D}
    \end{equation}
with scalar parameters $A$, $B_p$, and $B_d$. The exponential term $e^{A (t - \tau)}$ renders the state dependent on the specific power trajectory and not only on the total energy consumption.
For instance, early power consumption may lead to higher state-dependent losses and, consequently, to a lower terminal state than late power consumption. 

Leveraging the exponential term of (\ref{eq:analyticalSolution1D}), we define TI envelopes by weighting the power inputs by the worst-case time-varying coefficients, i.e., the smallest weight for the lower energy bound and the largest weight for the upper energy bound, to provide compliant energy boundaries independent from when power is consumed. 
We further elaborate on this idea in this section and in Section~\ref{sec:operationalPotentialMultiD} to define TI energy envelopes.

\subsection{TI Flexibility Envelope Definition}

We introduce the following theorem as a basis for the definition of TI energy flexibility envelopes: 

\begin{theorem}
    If a power consumption trajectory $p_{a}$ fulfills the power constraints and, for all $t \in \mathcal{T}$: 
    \begin{equation}
        \underbrace{\int_0^t e^{-A \tau} p_{-} (\tau) \text{d}\tau}_{\text{E}_{\text{down}}^{\text{\tiny TI}}(t)} \leq \int_0^t p_a(\tau) \text{d} \tau \leq \underbrace{\int_0^t e^{A (t-\tau)} p_{+} (\tau) \text{d}\tau}_{\text{E}_{\text{up}}^{\text{\tiny TI}} (t)}, 
        \label{eq:opEnergyBound1D}
    \end{equation}
    where ${p}_{-}$ and ${p}_{+}$ describe feasible power consumption trajectories, i.e., fulfilling the power and state constraints, that are exposed to the same disturbances, i.e. heat gains $d$ and initial conditions $x_0$ as trajectory $a$, then, trajectory $a$ fulfills the state constraints, i.e.,
    \begin{equation}
        x_{\text{min}} \leq x_a (t) \leq x_{\text{max}}, \quad \forall t \in \mathcal{T},
    \end{equation}
    and $\text{E}_{\text{up}}^{\text{\tiny TI}}$ and $\text{E}_{\text{down}}^{\text{\tiny TI}}$ are TI energy flexibility bounds. 
\end{theorem}

\begin{IEEEproof}
    
    \textit{Upper Bound:} Let us consider two power consumption trajectories, ${p}_a$ and ${p}_{+}$, and their resulting states ${x}_a$ and ${x}_{+}$, respectively. 
    Since both trajectories start with the same initial state, $x_0$, and are exposed to the same disturbances, $d$, the states' difference at time $t$ is:
    \begin{equation}
        x_a (t) - x_{+} (t) =  B_p \int_0^t e^{A (t - \tau) } \left( p_a(\tau) - p_{+} (\tau)  \right) \text{d} \tau.
        \label{eq:stateDiff1D}
    \end{equation}
    Since the system is lossy, $A \leq 0$, and therefore:
    \begin{equation}
        e^{A (t-\tau)} \leq 1, \quad 0\leq \tau \leq t, \quad t \in \mathcal{T}. 
        \label{eq:opEnergyBound1D_step2}
    \end{equation}
    Hence, given the right inequality in (\ref{eq:opEnergyBound1D}) combined with (\ref{eq:opEnergyBound1D_step2}):
    \begin{equation}
        \int_0^t e^{A (t-\tau)} p_{a} (\tau) \text{d}\tau \leq \int_0^t e^{A (t-\tau)} p_{+} (\tau) \text{d}\tau.
    \end{equation}
    Consequently, using $B_p \geq 0$ in (\ref{eq:stateDiff1D}), we can guarantee that trajectory $a$ fulfills the upper state constraints, i.e.
    \begin{equation}
        x_a (t) \leq x_{+} (t) \leq x_{\text{max}}, \quad \forall t \in \mathcal{T}.
    \end{equation}
    
    \textit{Lower Bound:} Similarly, since $A \leq 0$, it holds that: 
    \begin{equation}
        e^{A (t-\tau)} \geq e^{A t}, \quad 0\leq \tau \leq t, \quad t \in \mathcal{T}. 
        \label{eq:exp_LB}
    \end{equation}
    Therefore, combining the left inequality in (\ref{eq:opEnergyBound1D}) with (\ref{eq:exp_LB}), we obtain: 
    \begin{equation}
        \int_0^t e^{A (t- \tau)} p_{-} (\tau) \text{d}\tau \leq \int_0^t e^{A (t-\tau)} p_a(\tau) \text{d} \tau.
    \end{equation}
    Consequently, as $B_p \geq 0$ and since ${p}_-$ satisfies the state constraints, we can guarantee that the lower state constraints are satisfied, i.e.: 
    \begin{equation}
        x_a (t) \geq x_{-} (t) \geq x_{\text{min}}, \quad \forall t \in \mathcal{T}.
    \end{equation}
\end{IEEEproof}

\begin{remark}
    The upper and lower energy bounds, respectively $\text{E}_{\text{up}}^{\text{\tiny TI}}$ and $\text{E}_{\text{down}}^{\text{\tiny TI}}$, correspond to the weighted sum of power values. Yet, grid operators utilizing flexibility only need to monitor the system's energy consumption, conveniently ignoring time-varying weights. 
\end{remark}

\subsection{Computation of TI Flexibility Bounds}

In practice, we aim to find the power consumption trajectories ${p}_{-}$ and ${p}_{+}$ that maximize the TI flexibility offered by the system, i.e., that maximize the width of the TI envelope formulated in Theorem 1. Hence, the TI upper energy bound can be computed as the optimal objective function's value of:
\begin{subequations}
\begin{align}
    \text{E}_{\text{up}}^{\text{\tiny TI}^*} (t)  = \max_{{p}_{+}} & \quad \int_0^t e^{A (t-\tau)} p_{+} (\tau) \text{d}\tau, \\
    & \text{s.t. Constraints (\ref{eq:upperBound_1})-(\ref{eq:upperBound_3})}. 
\end{align}  
\label{eq:upperBound_op}
\end{subequations}
By discretizing\footnote{The finite difference method is consistently used to discretize problems.} the above problem, we obtain a linear optimization problem that can be solved efficiently. The TI lower energy bound can be computed similarly. 

\begin{remark}[Connection to TD energy envelope]
    The TD energy envelope computed in Section~\ref{subsec:MaxEnvelopeDef} can lead to state constraint violations when used in operation. Nevertheless, by solving~(\ref{eq:upperBound}), we obtain a power trajectory that satisfies the thermal comfort constraints. Hence, this trajectory can be used to form exponentially-weighted TI bounds $\text{E}_{\text{up}}^{\text{\tiny TI}}$ and $\text{E}_{\text{down}}^{\text{\tiny TI}}$ according to (\ref{eq:opEnergyBound1D}). However, there is no guarantee that the resulting bounds would maximize the proposed TI envelope's width. 
\end{remark}

\section{Trajectory Independent (TI) Energy Flexibility: Multi-Dimensional Systems}
\label{sec:operationalPotentialMultiD}

In this section, we extend the proof provided for a uni-dimensional system to a multi-dimensional system, i.e., a system with multiple coupled states. Multi-dimensional flexible systems can be operated in a distributed or centralized manner. In a distributed setting, resources can offer their flexibility individually, i.e., resources are physically connected, but each offers flexibility independently, leading to as many flexibility envelopes as the number of resources. In a centralized setup, resources offer their total flexibility together, i.e., they are aggregated into one flexible entity. Consequently, only one energy flexibility envelope is used to describe their flexibility. This paper first offers an intuition of the additional challenges appearing in multi-dimensional systems and proposes formulations of TI flexibility envelopes for both configurations. 

\subsection{Intuition}

To develop TI envelopes in the uni-dimensional case, we exploit the idea of worst time-varying weights. However, in a multi-dimensional case, power inputs may also have different impacts on different system states. Hence, TI flexibility envelopes should only enclose power consumption paths that are acceptable for all system states. In this section, we extend the uni-dimensional intuition by considering not only the worst time-varying but also the worst state-dependent weights in multi-dimensional systems. 

\subsection{Flexible Resources in a Distributed Set-Up}
\label{subsec:energyDecoupled}

In a distributed setting, physically connected loads offer flexibility independently. Hence, each load's TI flexibility envelope should guarantee its constraints, irrespective of other loads' energy consumption. 

\subsubsection{TI Flexibility Envelope Definition}
We first introduce Theorem~\ref{th:decentralized_coupled} which defines bounds that constrain the energy consumption of the set of considered loads. Then, in Theorem~\ref{th:decentralized_decoupled}, we present a method that decouples the energy envelopes of the different loads. 

\begin{theorem}[Coupled Energy Envelopes]
    \label{th:decentralized_coupled}
     We assume that the energy consumption of multiple physically coupled loads, characterized by the power trajectory $\boldsymbol{p}_{a}$, belongs, for all $t \in \mathcal{T}$, to the polytope: 
    \begin{equation}
    \left\{
        \begin{aligned}
            & \boldsymbol{\alpha}(t) \int_0^t \boldsymbol{p}_a (\tau ) \text{d} \tau \leq \boldsymbol{b}_{+} (t), \\
            & \boldsymbol{\beta}(t) \int_0^t \boldsymbol{p}_a (\tau ) \text{d} \tau \geq \boldsymbol{b}_{-} (t) , 
        \end{aligned}
    \right.
    \label{eq:polytope}
    \end{equation}
    where the matrices $\boldsymbol{\alpha}(t)$ and $\boldsymbol{\beta}(t)$ are defined component-wise as: 
    \begin{equation}
        \alpha_{i,j}(t) = \max_{0 \leq \tau \leq t} \left( e^{\mathbf{A} (t - \tau)} \mathbf{B}_p\right)_{i,j}, 
        \label{ref:defAlpha}
    \end{equation}
    \begin{equation}
        \beta_{i,j}(t) = \min_{0 \leq \tau \leq t} \left( e^{\mathbf{A} (t - \tau)} \mathbf{B}_p\right)_{i,j}, 
    \end{equation}
    and the vectors $\boldsymbol{b}_{+} (t)$ and $\boldsymbol{b}_{-} (t)$ are: 
    \begin{equation}
        \boldsymbol{b}_{+} (t) = \int_0^t e^{\mathbf{A} (t - \tau)} \mathbf{B}_p \hspace{0.1cm} \boldsymbol{p}_+ (\tau ) \text{d} \tau,
    \end{equation}
    \begin{equation}
        \boldsymbol{b}_{-} (t) = \int_0^t e^{\mathbf{A} (t - \tau)} \mathbf{B}_p \hspace{0.1cm} \boldsymbol{p}_- (\tau ) \text{d} \tau,
    \end{equation} 
    where $\boldsymbol{p}_{-}$ and $\boldsymbol{p}_{+}$ describe feasible power consumption trajectories, i.e., fulfilling the power and state constraints, that are exposed to the same heat gains $\boldsymbol{d}$ and initial conditions $\boldsymbol{x}_0$ as trajectory $a$. Then, trajectory $a$ fulfills the state constraints.
\end{theorem}

\begin{IEEEproof}
    The analytical solution of (\ref{eq:subID}) in the multi-dimensional case is: 
    \begin{equation}
        \boldsymbol{x} (t) = e^{\mathbf{A} t} \boldsymbol{x}_0 + \int_0^t e^{\mathbf{A} (t - \tau)} \left( \mathbf{B}_d \boldsymbol{d} (\tau) + \mathbf{B}_p \boldsymbol{p} (\tau) \right) \text{d}\tau.
    \end{equation}
    
    \textit{Upper Bound:} Let us consider two power consumption trajectories $\boldsymbol{p}_a$ and $\boldsymbol{p}_+$ and their resulting states $\boldsymbol{x}_a$ and $\boldsymbol{x}_+$, respectively. At time instant $t$, the states' difference of the $i^{\text{th}}$ load is given by: 
    \begin{equation}
    \begin{aligned}
         x_{a,i} & (t) -  x_{+,i} (t) \\ 
         & = \int_0^t  \left( e^{\mathbf{A} (t - \tau)} \mathbf{B}_p \left( \boldsymbol{p}_{a}(\tau) - \boldsymbol{p}_{+} (\tau) \right) \right)_i  \text{d}\tau, \\
         & = \int_0^t \sum_{j \in \mathcal{I}} \left( e^{\mathbf{A} (t - \tau)} \mathbf{B}_p\right)_{i,j} \left( p_{a,j} - p_{+,j} \right) (\tau) \text{d}\tau. 
    \end{aligned}
    \label{eq:differenceStateMD1}
    \end{equation}
    According to (\ref{eq:polytope}), the energy consumed in trajectory $a$ fulfills: 
    \begin{equation}
        \boldsymbol{\alpha}(t) \int_0^t \boldsymbol{p}_a (\tau ) \text{d} \tau \leq \underbrace{\int_0^t e^{\mathbf{A} (t - \tau)} \mathbf{B}_p \hspace{0.1cm} \boldsymbol{p}_+ (\tau ) \text{d} \tau}_{\boldsymbol{b}_{+} (t)},
        \label{eq:systemUp}
    \end{equation}
    i.e., the following inequality holds for the $i^{\text{th}}$ load: 
    \begin{equation}
       \hspace{-0.3cm} \sum_{j \in \mathcal{I}} \alpha_{i,j}(t) \int_0^t {p}_{a,j} (\tau ) \text{d} \tau \leq  \int_0^t \sum_{j \in \mathcal{I}} \left( e^{\mathbf{A} (t - \tau)} \mathbf{B}_p\right)_{i,j} p_{+,j}  (\tau) \text{d}\tau.
        \label{eq:systemUp}
    \end{equation}
    Therefore, by combining the definition of $\boldsymbol{\alpha} (t)$ in (\ref{ref:defAlpha}) and (\ref{eq:systemUp}), we obtain: 
    \begin{equation}
    \begin{aligned}
        \int_0^t & \sum_{j \in \mathcal{I}} \left( e^{\mathbf{A} (t - \tau)} \mathbf{B}_p\right)_{i,j} p_{a,j}  (\tau) \text{d}\tau \\ 
        & \leq \sum_{j \in \mathcal{I}} \alpha_{i,j}(t) \int_0^t {p}_{a,j} (\tau ) \text{d} \tau \\
        & \leq \int_0^t \sum_{j \in \mathcal{I}} \left( e^{\mathbf{A} (t - \tau)} \mathbf{B}_p\right)_{i,j} p_{+,j}  (\tau) \text{d}\tau.
    \end{aligned}
    \end{equation}
    According to (\ref{eq:differenceStateMD1}), this implies that: 
    \begin{equation}
        x_{a,i} (t) \leq x_{+,i} (t), \quad \forall i \in \mathcal{I}, \quad \forall t \in \mathcal{T}. 
    \end{equation}
    
    \textit{Lower Bound:} A similar reasoning can be employed to derive the lower bound state constraint.
\end{IEEEproof}
\begin{remark}
    In the multi-dimensional case, it is not possible to directly leverage the inequalities used in (\ref{eq:opEnergyBound1D_step2}) and (\ref{eq:exp_LB}), as $e^{\mathbf{A} (t - \tau)}\mathbf{B}_p$ is not necessarily component-wise minimum at $\tau=0$ and maximum at $\tau=t$. Therefore, we define $\boldsymbol{\alpha}$ and $\boldsymbol{\beta}$ explicitly selecting the maxima and minima components.
\end{remark}

Theorem~\ref{th:decentralized_coupled} results in a linear system of inequalities. Therefore, the different flexible loads must coordinate to fulfill the set of inequalities. However, in a distributed setup, each load decides upon its flexibility offer without coordinating with others. Hence, we introduce Theorem~3, which aims to determine the largest box inscribed in the polytope (\ref{eq:polytope}) to decouple the energy constraints of the different loads. 

\begin{theorem}
\label{th:decentralized_decoupled}
    We define, for all $t \in \mathcal{T}$, the lower and upper TI energy bounds as: 
    \begin{equation}
    \begin{aligned}
        \left( \mathbf{E}_{\text{up}}^{\text{\tiny TI} \text{,d}}, \mathbf{E}_{\text{down}}^{\text{\tiny TI}\text{,d}} \right) (t) = \argmax_{\mathbf{E}_+, \mathbf{E}_-} & \hspace{0.1cm} \sum_{i \in \mathcal{I}} \ln \left( \text{E}_{+,i} - \text{E}_{-,i} \right)  \\
        \text{s.t. } \hspace{0.2cm} & \boldsymbol{\alpha}(t) \mathbf{E}_+  \leq \boldsymbol{b}_{+} (t), \\
        & \boldsymbol{\beta}(t) \mathbf{E}_-  \geq \boldsymbol{b}_{-} (t), \\
         & \mathbf{E}_+  \geq \mathbf{E}_-,
    \end{aligned}
    \label{eq:largestBox}
    \end{equation}
    describing the largest box inscribed in the polytope (\ref{eq:polytope}), with $\boldsymbol{\alpha}$, $\boldsymbol{\beta}$, $\boldsymbol{b}_+$ and $\boldsymbol{b}_-$ defined according to Theorem 2.
    If the power consumption trajectory of every load $i$ fulfills: 
    \begin{equation}
        \text{E}_{\text{down},i}^{\text{\tiny TI}\text{,d}} (t) \leq \int_0^t p_{a,i} (\tau) \text{d}\tau \leq \text{E}_{\text{up},i}^{\text{\tiny TI}\text{,d}} (t),  \quad \forall t \in \mathcal{T},
    \end{equation}
    then trajectory $a$ fulfills the state constraints.
\end{theorem}

\begin{IEEEproof}
    First, we explain why the solution of (\ref{eq:largestBox}) describes the largest box, i.e., the box with the maximum area, inscribed in the polytope (\ref{eq:polytope}). 
    
    \textit{Largest Box in a Polytope \cite{BEMPORAD2004151}:} The largest box inscribed in a polytope $\mathcal{P} = \{ \boldsymbol{x} \in \mathbb{R}^{N}, \hspace{0.1cm} \boldsymbol{\Phi} \boldsymbol{x} \leq \boldsymbol{\phi} \}$ can be described by its vertices, which must all belong to the polytope. More specifically, each vertex of the box can be described as $\boldsymbol{x} + \mathbf{V}(S) \boldsymbol{y}$, where $\boldsymbol{x}$ characterizes the position of the box and $\boldsymbol{y}$ the length of the edges. For any subset  $S \subseteq \{1, \cdots, N\}$, $\mathbf{V}(S)$ is an indicator matrix, i.e., its diagonal coefficients are equal to 1 if $j \in S$ and 0 otherwise. The largest inscribed box problem can be formulated as: 
    \begin{equation}
        \begin{aligned}
            \max_{\boldsymbol{x}, \boldsymbol{y}} \hspace{0.2cm} & \prod_{i =1}^N  y_i  \\
            \text{s.t. } \hspace{0.2cm} & \boldsymbol{\Phi} \boldsymbol{x} + \boldsymbol{\Phi} \mathbf{V}(S) \boldsymbol{y} \leq \boldsymbol{\phi}, \hspace{0.3cm} \forall S \in \mathbb{P} \left( \{1, \cdots, N\} \right) \\
            & \boldsymbol{y} \geq 0,
            \label{eq:polytopeDemPbIntital}
        \end{aligned}
    \end{equation}
    where $\mathbb{P} \left(X \right) $ is the power set of $X$, i.e., the set of all subsets of $X$.
    The first constraint can be written componentwise as: 
    \begin{equation}
        \sum_{j=1}^N \Phi_{i,j} x_j + \sum_{j \in S} \Phi_{i,j} y_j \leq \phi_i, \hspace{0.0cm} \forall S \in \mathbb{P} \left( \{1, \cdots, N\} \right).
        \label{eq:polytopeDem}
    \end{equation}
    Let us consider the subset $S_i^+ \in \mathbb{P} \left( \{1, \cdots, N\} \right)$ such that $j$ belongs to $S_i^+$ if and only if $\Phi_{i,j} \geq 0$. As $\boldsymbol{y} \geq 0$, applying (\ref{eq:polytopeDem}) to $S_i^+$ yields: 
    \begin{equation}
        \sum_{j=1}^N \Phi_{i,j} x_j + \sum_{j \in S} \Phi_{i,j} y_j \leq \sum_{j=1}^N \Phi_{i,j} x_j + \sum_{j \in S_i^+} \Phi_{i,j} y_j \leq \phi_i,
        \label{eq:polytopeDem2}
    \end{equation}
    for all $S \in \mathbb{P} \left( \{1, \cdots, N\} \right)$. This demonstrates that the satisfaction of (\ref{eq:polytopeDem}) applied to subset $S_i^+$ implies the satisfaction of (\ref{eq:polytopeDem}) for all subsets of $\{1, \cdots, N\}$. Hence, problem (\ref{eq:polytopeDemPbIntital}) can be written as: 
    \begin{equation}
        \begin{aligned}
            \max_{\boldsymbol{x}, \boldsymbol{y}} \hspace{0.2cm} & \sum_{i =1}^N  \ln \left(y_i \right)  \\
            \text{s.t. } \hspace{0.2cm} & \boldsymbol{\Phi} \boldsymbol{x} + \boldsymbol{\Phi}^p \boldsymbol{y} \leq \boldsymbol{\phi}, \\
            & \boldsymbol{y} \geq 0. 
            \label{eq:polytopeDemPbIntital2}
        \end{aligned}
    \end{equation}
    where the transformation $\mathbf{X}^{p}$ of matrix $\mathbf{X}$, applied here to $\boldsymbol{\Phi}$, is defined component-wise as $X^{p}_{i,j} = \max \left(X_{i,j}, 0\right)$. For further details, we refer the interested reader to \cite{BEMPORAD2004151}.
    
    \textit{Largest Box in (\ref{eq:polytope}):} By setting $\boldsymbol{x} = \mathbf{E}_-$ and $\boldsymbol{y} = \mathbf{E}_+ - \mathbf{E}_-$, we obtain: 
    \begin{equation}
    \begin{aligned}
        \hspace{-0.1cm}\max_{\mathbf{E}_+, \mathbf{E}_-} & \hspace{0.1cm} \sum_{i \in \mathcal{I}} \ln \left( \text{E}_{+,i} - \text{E}_{-,i} \right)  \\
        \text{s.t. } \hspace{0.2cm} & \boldsymbol{\alpha}(t) \mathbf{E}_- + \boldsymbol{\alpha}(t)^p \left( \mathbf{E}_+ - \mathbf{E}_- \right) \leq \boldsymbol{b}_{+} (t), \\
        & -\boldsymbol{\beta}(t) \mathbf{E}_- + \left(-\boldsymbol{\beta}(t) \right)^p \left( \mathbf{E}_+ - \mathbf{E}_- \right) \leq -\boldsymbol{b}_{-} (t), \\
        & \mathbf{E}_+ - \mathbf{E}_- \geq 0.
    \end{aligned}
    \label{eq:largestBoxProof}
    \end{equation}
    As we assume the off-diagonal elements of $\mathbf{A}$ to be non-negative, so are the ones of $\mathbf{A} (t-\tau)$ for $\tau \leq t$. Hence, the components of $e^{\mathbf{A} (t-\tau)}$ are non-negative \cite{DELEENHEER2001259}. Then, as the components of $\mathbf{B}_p$ are all non-negative, $\boldsymbol{\gamma}_{+}$ is non-negative. Therefore, all components of $\boldsymbol{\alpha}(t)$ and $\boldsymbol{\beta}(t)$ are non-negative, which finally leads to (\ref{eq:largestBox}).
    
    \textit{State Constraint Satisfaction:} Finally, as an inscribed box is an inner approximation of a polytope, any energy consumption value in the box also belongs to the polytope. Consequently, it satisfies the state constraints, according to Theorem 2. 
\end{IEEEproof}

\subsubsection{Computation of TI Flexibility Bounds}
In practice, the goal is to determine the TI flexibility bounds that maximize the total available flexibility as formulated in Theorem~3. Hence, we need to find power consumption trajectories, $\boldsymbol{p}_+$ and $\boldsymbol{p}_-$, that maximize the inscribed box's area while satisfying the power and state constraints. 

After discretizing the problem, we observe that the bounds $\boldsymbol{b}_{+,k}$ and $\boldsymbol{b}_{-,k}$ are linear combinations of the discretized power vectors $\boldsymbol{p}_+$ and $\boldsymbol{p}_-$:
\begin{equation}
    \boldsymbol{b}_{+,k} = \Delta t \sum_{l \in \mathcal{T}_d} e^{\mathbf{A} (k-l)} \mathbf{B}_p \boldsymbol{p}_{+,l} = \boldsymbol{\mu}_{+,k}^\intercal \hspace{0.05cm} \boldsymbol{p}_+,
\end{equation}
\begin{equation}
    \boldsymbol{b}_{-,k} = \Delta t \sum_{l \in \mathcal{T}_d} e^{\mathbf{A} (k-l)} \mathbf{B}_p \boldsymbol{p}_{-,l} = \boldsymbol{\mu}_{-,k}^\intercal \hspace{0.05cm} \boldsymbol{p}_-,
\end{equation}
where $\Delta t$ denotes the discretization step. Hence, the overall problem can be formulated as the following convex optimization: 
\begin{equation}
\hspace{-0cm}
    \begin{aligned}
        & \hspace{-1cm} \max_{\mathbf{E}_+, \mathbf{E}_-, \boldsymbol{p}_+, \boldsymbol{p}_-} \hspace{0.2cm} \sum_{k \in \mathcal{T}_d} \sum_{i \in \mathcal{I}} \ln \left( \text{E}_{+,k,i} - \text{E}_{-,k,i} \right) \\
        \text{s.t. } \hspace{0.2cm} & \boldsymbol{\alpha}_k \mathbf{E}_{+,k}  \leq \boldsymbol{\mu}_{+,k}^\intercal \hspace{0.05cm} \boldsymbol{p}_+, \hspace{2.6cm} \forall k \in \mathcal{T}_d, \\
        & \boldsymbol{\beta}_k \mathbf{E}_{-,k} \geq  \boldsymbol{\mu}_{-,k}^\intercal \hspace{0.05cm}\boldsymbol{p}_-, \hspace{2.6cm} \forall k \in \mathcal{T}_d,\\
        & \text{(\ref{eq:upperBound_1})} - \text{(\ref{eq:upperBound_3})} \text{ \small in the discretized form applied to } \boldsymbol{p}_-, \\
        & \text{(\ref{eq:upperBound_1})} - \text{(\ref{eq:upperBound_3})} \text{ \small in the discretized form applied to }  \boldsymbol{p}_+. \\
        & \mathbf{E}_+ \geq \mathbf{E}_-.
    \end{aligned}
    \label{eq:optimalBox}
\end{equation}

\begin{figure}
    \hspace{0.4cm}
    \includegraphics[width = 0.7\columnwidth]{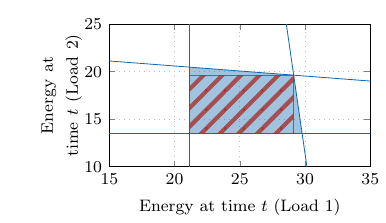}
    \caption{Largest 2-dimensional box, represented in striped red, inside a polytope, represented in blue. The blue lines describe the limits of the polytope.}
    \label{fig:rectanglePolytope}
\end{figure}

\begin{remark}
Theorem~2 first defines the worst time-varying weights, $\boldsymbol{\alpha}$ and $\boldsymbol{\beta}$. Then, Theorem~3 states that the energy consumption of a load must be robust to the worst-case energy consumption of other loads.  Fig.~\ref{fig:rectanglePolytope} provides a 2-dimensional example of the problem for one timestep. The blue area designates the polytope described in~(\ref{eq:polytope}), formed using the optimal power trajectory obtained from (\ref{eq:optimalBox}) to determine $\boldsymbol{b}_+$ and $\boldsymbol{b}_-$. The striped red area represents the largest rectangle included in the polytope, determined by the optimal energy value in (\ref{eq:optimalBox}). The red lines delimiting the rectangle define the upper and lower TI energy bounds for the individual loads, which are decoupled.
\end{remark}

\subsection{Flexible Loads in a Centralized Set-Up}
\label{subsec:coupledFormulation}

The formulation presented in Section~\ref{subsec:energyDecoupled} may become conservative in cases where flexible loads are strongly coupled. Indeed, to offer individual TI energy flexibility envelopes, the previous formulation assumes the worst-case power exchanges between loads. Therefore, we suggest a second TI flexibility envelope formulation where the loads form a flexibility pool that is centrally managed. In such a case, we define TI energy flexibility bounds for the total energy consumed by the pool, as opposed to the previous individual energy constraints. 

When resources are managed centrally, a power flexibility request sent to the pool is distributed among loads according to a central plan. Specifically, we can assume that the dispatch plan, i.e., the way flexibility is dispatched among loads, is fixed and is independent of the amount of flexibility to be dispatched. In this case, given a total power consumption request ${p}_{\text{tot}}$, the load $i$ consumes: 
\begin{equation}
    \label{eq:dispatch}
    p_{i} (t) = \delta_i(t) p_{\text{tot}} (t), \quad \forall t \in \mathcal{T}.
\end{equation}
The vector containing the dispatch factors $\boldsymbol{\delta}$ can vary over time, but, at all times, the aggregated power of individual loads must match the total requested power, i.e.: 
\begin{equation}
    \sum_{i \in \mathcal{I}} \delta_i (t) = 1, \quad \forall t \in \mathcal{T}.
\end{equation}

\subsubsection{TI Flexibility Envelope Definition}

In a distributed setting, TI envelopes must be robust to power consumption trajectories of other loads, as all loads are operated independently. In a centralized setting, flexible consumers operate as a group, leading to a less conservative assumption of the behavior of surrounding loads compared to the distributed setup. The centralized TI envelope formulation leverages this information. 
\begin{theorem}
    It is assumed that the total power consumption trajectory of a pool of flexible consumers, $p_{a,\text{tot}}$, satisfies individual power constraints and is contained in an energy envelope delimited, for all $t \in \mathcal{T}$, by: 
    \begin{equation}
        \begin{aligned}
            \hspace{-0.49cm}\text{E}_{\text{up}}^{\text{\tiny TI}\text{,c}} (t) & = \min_{i \in \mathcal{I}} \left[ \frac{1}{\gamma_{+,i} (t)} \left( \int_0^t e^{\mathbf{A} (t - \tau)} \mathbf{B}_p \boldsymbol{p}_{+} ( \tau) \text{d} \tau \right)_{i} \right], \\
            \hspace{-0.49cm}\text{E}_{\text{down}}^{\text{\tiny TI}\text{,c}} (t) & = \max_{i \in \mathcal{I}} \left[ \frac{1}{\gamma_{-,i} (t)} \left( \int_0^t e^{\mathbf{A} (t - \tau)} \mathbf{B}_p \boldsymbol{p}_{-} ( \tau) \text{d} \tau \right)_{i} \right],
        \end{aligned}
        \label{eq:bounds_th4}
    \end{equation}
    where $\boldsymbol{\gamma}_{+} (t)$ and $\boldsymbol{\gamma}_{-} (t)$ are defined component-wise as: 
    \begin{equation}
    \begin{aligned}
        \gamma_{+,i} (t) & = \max_{0 \leq \tau \leq t} \sum_{j \in \mathcal{I}} \left( e^{\mathbf{A} (t - \tau)} \mathbf{B}_p\right)_{i,j} \delta_j (\tau), \\
        \gamma_{-,i} (t) & = \min_{0 \leq \tau \leq t} \sum_{j \in \mathcal{I}} \left( e^{\mathbf{A} (t - \tau)} \mathbf{B}_p\right)_{i,j} \delta_j (\tau),
    \end{aligned}
    \label{eq:defGamma}
    \end{equation}
    and $\boldsymbol{p}_+$ and $\boldsymbol{p}_-$ describe feasible power consumption trajectories, i.e., fulfilling the power and state constraints, dispatched according to $\boldsymbol{\delta}$, and that are exposed to the same heat gains $\boldsymbol{d}$ and initial conditions $\boldsymbol{x}_0$ as trajectory $a$. Then, trajectory $a$ fulfills the state constraints.
\end{theorem}

\begin{IEEEproof}
    We use similar arguments as in Theorem 2, additionally including the fixed dispatch plan $\boldsymbol{\delta}$.
    
    \textit{Upper Bound:} We consider two total power consumption trajectories, ${p}_{a,\text{tot}}$ and ${p}_{+,\text{tot}}$. Since both total power trajectories are dispatched according to $\boldsymbol{\delta}$, the difference between the $i^{\text{th}}$ state of both trajectories, at time instant $t$, is:
    \begin{equation}
    \begin{aligned}
         & x_{a,i}  (t) -  x_{+,i} (t) \\ 
         & = \int_0^t \sum_{j \in \mathcal{I}} \left( e^{\mathbf{A} (t - \tau)} \mathbf{B}_p\right)_{i,j} \delta_j(\tau) \left( p_{a,\text{tot}} - p_{+,\text{tot}} \right) (\tau) \text{d}\tau. 
    \end{aligned}
    \label{eq:differenceStaetMD1}
    \end{equation}
    As the total energy consumption of trajectory $a$ is upper-bounded by $\text{E}_{\text{up}}^{\text{\tiny TI}\text{,c}} (t)$ at time instant $t$, we can state, for all $i \in \mathcal{I}$, that: 
    \begin{equation}
    \begin{aligned}
        \int_0^t & p_{a,\text{tot}} (\tau) \text{d} \tau \text{ } \leq \text{ }\text{E}_{\text{up}}^{\text{\tiny TI}\text{,c}} (t) \\
        & \leq \frac{1}{\gamma_{+,i} (t)} \int_0^t \sum_{j \in \mathcal{I}} \left( e^{\mathbf{A} (t - \tau)} \mathbf{B}_p\right)_{i,j} p_{+, j} ( \tau) \text{d} \tau.
        \label{eq:intermediateStepTh4}
    \end{aligned}
    \end{equation}
    As we assume the off-diagonal elements of $\mathbf{A}$ and all the components of $\mathbf{B}_p$ to be non-negative, $\boldsymbol{\gamma}_{+}$ is non-negative.
    By combining the definition of $\boldsymbol{\gamma}_{+}$ given in (\ref{eq:defGamma}) and (\ref{eq:intermediateStepTh4}), we obtain: 
    \begin{equation}
    \begin{aligned}
        \int_0^t \sum_{j \in \mathcal{I}} & \left( e^{\mathbf{A} (t - \tau)} \mathbf{B}_p\right)_{i,j} \overbrace{\delta_j (\tau) p_{a,\text{tot}} (\tau)}^{p_{a,j} (\tau)} \text{d} \tau \\
        & \leq \gamma_{+,i} (t) \int_0^t p_{a,\text{tot}} (\tau) \text{d} \tau \\
        & \leq \int_0^t \sum_{j \in \mathcal{I}} \left( e^{\mathbf{A} (t - \tau)} \mathbf{B}_p\right)_{i,j} p_{+, j} ( \tau) \text{d} \tau.
    \end{aligned}
    \end{equation}
    Therefore, according to (\ref{eq:differenceStaetMD1}), we can guarantee that trajectory $a$ satisfies the state constraints.

    \textit{Lower Bound:}  A similar reasoning can be implemented to obtain the lower bound state constraint.
\end{IEEEproof}

\begin{remark}
    For this centralized setup, we first must ensure that any power consumption trajectories are feasible, regardless of when power is consumed. Then, we must ensure that power trajectories lead to acceptable state values, for all states. Therefore, we apply to power trajectories the worst time-varying weight ($\boldsymbol{\gamma}$) for all states, and we restrict energy consumption based on the worst bounds among all states. 
\end{remark}

\subsubsection{Computation of TI Flexibility Bounds}
To maximize the available flexibility, we aim to find the feasible power consumption trajectories $\boldsymbol{p}_+$ and $\boldsymbol{p}_-$ which maximize the width of the TI flexibility envelope as formulated in Theorem 4. To this aim, we compute the upper energy flexibility bound as the optimal objective function's value of:
\begin{align}
     & \max_{\boldsymbol{p}_+} \quad \Bar{\text{E}} (t), \\
     & \text{s.t. } \text{Constraints (\ref{eq:upperBound_1})-(\ref{eq:upperBound_3})}, \\
    \hspace{-0.5cm}  & \Bar{\text{E}} (t) \leq \frac{1}{\gamma_{+,i} (t)} \left( \int_0^t e^{\mathbf{A} (t - \tau)} \mathbf{B}_p \boldsymbol{p}_{+} ( \tau) \text{d} \tau \right)_{i}, \forall i \in \mathcal{I}.
\end{align}  
The operational lower energy bound can be computed similarly.

\section{Case Study: Building Thermal Dynamics}
\label{sec:caseStudy}

This section first introduces a general linear model to describe a building's thermal dynamics and then describes the buildings later used as case studies.
The heating system of a building is a flexible load as it can shift its energy consumption over time while satisfying the heating system's power constraints and the inhabitants' thermal comfort constraints. Furthermore, buildings' thermal dynamics suffer from state-dependent losses. Therefore, we use this example to analyze the performance of the proposed formulations. 

\subsection{Linear Building Thermal Dynamics}

Thermal dynamics in a building can be approximated with a resistance-capacitance (RC) model, which consists of a simplified linear representation of the temperature evolution of different rooms in a building \cite{FINCK2019}. It can be written as: 
\begin{equation}
\begin{aligned}
    C_i \dfrac{dT_i}{dt} = & - \frac{T_i (t) - T_{a} (t)}{R_i}
     - \sum_{j \in \mathcal{A}_i} \frac{T_i (t) - T_j (t)}{R_{i,j}} \\
    & + p_{\text{th},i} (t) + d_{i} (t).
    \label{eq:generalBuilding}
\end{aligned}
\end{equation}
$T_i$ denotes the temperature of room $i$ that is connected to adjacent rooms, contained in the set of rooms $\mathcal{A}_i$, and $T_{a}$ is the ambient temperature. Variable $p_{\text{th},i}$ describes the thermal heating power input in room $i$, while $d_{i}$ represents all other sources of heat gains in room $i$. In the rest of the paper, we neglect additional heat gains due to occupancy, equipment and lighting for simplicity. Each room $i$ is characterized by its heat capacity, $C_i$, i.e., its ability to store heat, and its thermal resistances to the ambient environment and other adjacent rooms, $R_i$ and $\left( R_{i,j} \right)_{j \in \mathcal{A}_i}$ respectively. Additionally, we assume that each room is heated up by only one heating source and that each source directly influences only the room in which it is located.  

\subsection{Buildings}
\label{subsec:buildings}
To assess the proposed formulations, we introduce three examples: a one-room building, \textit{SwissHouse}, used as an illustrative example; Swiss building archetypes of one-zone equivalents of buildings, useful to understand the impact of building parameters on the performance of the proposed approach; and a 9-zone building, based on which we assess the multi-dimensional formulations. In all buildings, we assume to observe the initial conditions and thermal comfort range as stated in Table~\ref{tab:exampleOneZone}. Additionally, all buildings are exposed to the same ambient temperature, taken as the ambient temperature measured in Z\"urich, Switzerland between January, 5$^{\text{th}}$ and February, 5$^{\text{th}}$, 2020, which varies between -5°C and 15°C.
All envelopes are calculated for every day of this period starting at midnight for a horizon of one day.

\subsubsection{SwissHouse}
\label{subsubsec:swisshouse}
The one-zone \textit{SwissHouse} building is a representative building composed of one room. This building describes a light and well-insulated house, subject to an average Swiss climate \cite{energym}. The building's thermal parameters are detailed in Table~\ref{tab:exampleOneZone}.  

\subsubsection{Swiss Building Archetypes}
\label{subsubsec:archetype}
Building archetypes are examples of buildings, representative of the Swiss building stock. Since the formulation proposed in this paper heavily depends on the system's parameters, we study the formulation's performance for different one-zone building archetypes. As presented in \cite{guo2025}, we define 12~archetypes, each characterized by a combination of heat capacity and thermal resistance. More specifically, 3~different heat capacity values reflecting light, medium, and heavy construction, while 4 thermal resistances corresponding to different years of construction are defined. The parameter values are detailed in Table~\ref{tab:archetypeParameters}. For further details, we refer the interested reader to \cite{guo2025}.

\begin{table}[t]
  \centering
  \footnotesize
  \renewcommand{\arraystretch}{1.3}
  \setlength{\tabcolsep}{3pt}
  \caption{Swiss archetypes' parameter values.}
  \label{tab:archetypeParameters}
  \begin{tabular}{c cccc}
  \toprule
    \textbf{Parameter} & Unit & Type / Building Age & Values \\ \midrule
    $C$ & $\text{MJ} / \text{m}^2\text{K}$ & \begin{tabular}{c} Light \\
    Medium \\
    Heavy
    \end{tabular} & \begin{tabular}{c cccc} 0.1 \\
    0.3 \\
    0.5
    \end{tabular} \\\midrule
    $\dfrac{1}{R}$ & $\text{W} / \text{m}^2\text{K}$ & \begin{tabular}{c} Very well insulated ($>$ 2010) \\
    Well insulated (2000 - 2010) \\
    Medium insulated (1980-2000) \\
    Poorly insulated ($<$ 1980)
    \end{tabular} & \begin{tabular}{c cccc} 0.34 \\
    0.86 \\
    1.14 \\
    1.71
    \end{tabular} \\ \bottomrule
  \end{tabular}
\end{table}

\subsubsection{A 9-room building}
\label{subsec:caseStudyMultiD}

In \cite{BELAZI2022}, the thermal parameters of a 9-room building located in Montlu\c con, France are given. It comprises three floors, with three rooms on each floor, each of 20 m$^{\text{2}}$. The building is equipped with a heat pump characterized by an electric power rating of 9~kW. Each room can be individually controlled, with an individual power rating of 1~kW. The building is insulated from the outside with insulation panels. However, the rooms are poorly insulated from one another. To mimic individual independent apartments, the authors in \cite{BELAZI2022} additionally suggest installing supplementary insulation layers on indoor walls and floors. Both configurations, i.e., with and without supplementary indoor insulation, are used to assess the performances of the multi-dimensional formulations.

\section{Results}
\label{sec:results}

TI energy flexibility envelopes, presented in Sections~\ref{sec:operationalPotential1D} and~\ref{sec:operationalPotentialMultiD}, guarantee the satisfaction of state constraints in operation, e.g., inhabitants' thermal comfort in the case of flexible heating systems. In comparison, TD envelopes, presented in Section~\ref{sec:MaxPotential}, cannot guarantee the feasibility of all power consumption trajectories but describe the maximum and minimum energy values that a load can consume. 
In this section, we use \textit{Swisshouse} to define appropriate metrics, which will then be used to compare the formulations' performances on the \textit{Swiss Building Archetypes} and the \textit{9-room building}.

\subsection{Illustrative Example and Metrics}

\begin{figure}
    \hspace{0.7cm}
    \includegraphics[width = 0.73\columnwidth]{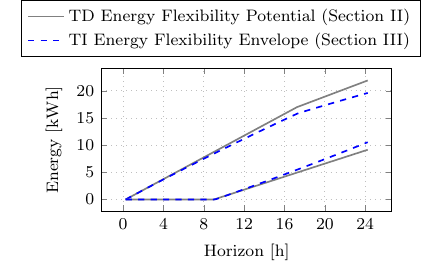}
    \caption{Example of the TI and TD flexibility envelopes of a one-zone building (\textit{SwissHouse}).}
    \label{fig:swisshouseEnvelopes}
\end{figure}

Fig.~\ref{fig:swisshouseEnvelopes} compares the TD envelope and the TI envelope of the one-zone building (\textit{SwissHouse}) over a time horizon of a day. In order to be robust against all power consumption trajectories, the TI lower energy bound gradually becomes larger than the TD lower energy bound. Similarly, the upper TI energy bound becomes smaller than the TD upper bound as the horizon increases. Overall, the area of the TI envelope decreases compared to the TD envelope's area. Hence, the first metric assesses the flexibility reduction between the TI and the TD envelopes, defined as the area between the envelope's bounds for different lead times. 

Furthermore, the TI upper bound may become smaller than the TI lower bound. This describes the \ac{MFPH}, our second metric. Past the \ac{MFPH}, for any energy consumption, there exists a power consumption trajectory that violates thermal comfort.

Additionally, all power consumption trajectories in TI flexibility envelopes satisfy the state constraints, while some trajectories in TD envelopes may violate them. Hence, our third metric assesses the worst thermal discomfort resulting from the use of TD envelopes, i.e., the worst absolute temperature deviation from the thermal comfort bounds. Integrating (\ref{eq:oneZoneExmaple}) reveals that consuming as much power as late as possible results in the largest final temperature, while early heating followed by the lowest possible power consumption results in the lowest final temperature. Hence, these power consumption trajectories are used to derive the worst thermal discomfort.

\subsection{Uni-Dimensional Systems: Swiss Building Archetypes}

The uni-dimensional TI flexibility formulation developed in Section~\ref{sec:operationalPotential1D} highly depends on the system's parameters. In the cases of buildings' thermal dynamics, these parameters are the thermal resistance and capacitance of a building. Therefore, we investigate the difference between the TI and TD uni-dimensional formulations for different building archetypes, representative of the Swiss building stock. 

\begin{figure}
    \hspace{0.2cm}
    \includegraphics[width = 0.8\columnwidth]{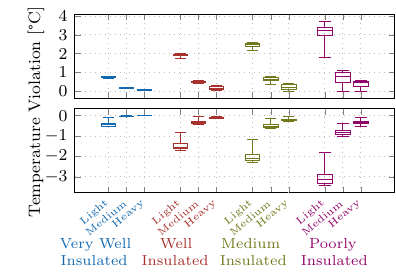}
    \caption{Worst thermal discomfort for different Swiss building archetypes, when consuming as much (upper plot) and as little energy as possible (lower plot) in the TD envelope.}
    \label{fig:boxplotParametersTemperatureViolation}
\end{figure}

Fig.~\ref{fig:boxplotParametersTemperatureViolation} displays the worst thermal discomfort experienced in different building archetypes. For highly insulated buildings or heavy constructions, the room temperature does not deviate by more than 0.5\textdegree C from the comfort bounds, which can be considered acceptable. However, inhabitants of poorly insulated buildings or light constructions may experience large thermal discomfort. For instance, in a poorly insulated building, the temperature may rise up to 3.8\textdegree C above the maximum acceptable temperature and sink 3.4\textdegree C below the minimum one. Inhabitants experiencing such thermal discomfort levels are likely to deregister from flexibility programs \cite{sweetnam2019}. 

\begin{figure}
    \hspace{-0cm}
    \includegraphics[width = 0.8\columnwidth]{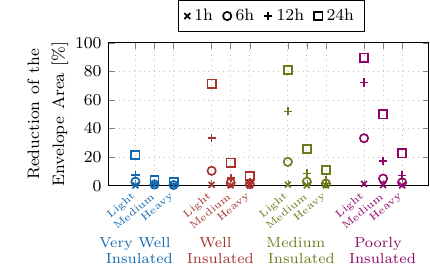}
    \caption{Reduction (median) of the flexibility region between the TI and the TD envelopes for different Swiss building archetypes and prediction horizons.}
    \label{fig:boxplotParametersFlexReductionTimesteps}
\end{figure}

In some archetypes, the discomfort that inhabitants may experience is large when employing TD envelopes in operation. Consequently, for the same archetypes, we expect TI envelopes to deviate significantly from TD envelopes. Fig.~\ref{fig:boxplotParametersFlexReductionTimesteps} describes the flexibility area reduction between the TI and the TD envelopes, for different time horizons, from one hour to one day. Combining Figs.~\ref{fig:boxplotParametersTemperatureViolation} and~\ref{fig:boxplotParametersFlexReductionTimesteps}, we observe that in the buildings that hardly experience discomfort when using TD envelopes, TI and TD envelopes do not differ significantly. However, the area of TI envelopes is significantly smaller than the area of TD envelopes for buildings affected by significant discomfort. Indeed, TI envelopes must ensure inhabitants' thermal comfort, even for the power consumption trajectory resulting in the worst thermal losses. As a consequence, in poorly insulated buildings, which suffer from large thermal losses, the TI flexibility available is smaller than the TD one. 

Fig.~\ref{fig:boxplotParametersFlexReductionTimesteps} also highlights the difference between the TI and TD formulations for different time horizons. One-hour ahead, the TI and TD flexibility envelopes have a similar width, for all archetypes. However, over longer prediction horizons, the area between TI energy envelopes becomes significantly smaller than the area between the TD envelopes, particularly for light constructions and poorly insulated buildings. For instance, one-day ahead, the TI flexibility envelope's area of a light, poorly insulated building is 90\% smaller than its TD envelope's area.

\begin{figure}
    \hspace{-1.7cm}
    \includegraphics[width = \columnwidth]{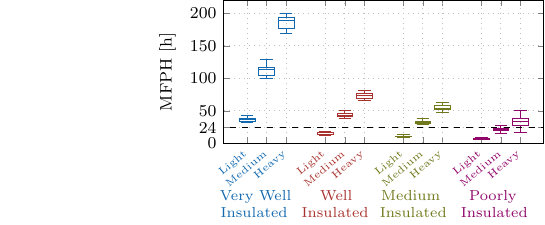}
    \caption{Maximum Flexibility Provision Horizon (MFPH) of the TI envelope for different Swiss building archetypes.}
    \label{fig:boxplotParametersMaxDuration}
\end{figure}

Some archetypes do not only offer a reduced flexibility envelope but also encounter a short \ac{MFPH}. Indeed, as Fig.~\ref{fig:boxplotParametersMaxDuration} indicates, the TI flexibility envelope of some archetypes is limited to a few hours. 
Aligned with previous observations, light constructions and/or poorly insulated buildings suffer from a small \ac{MFPH}. In light constructions, little thermal energy can be stored, and, in poorly insulated buildings, thermal losses to the ambient environment are very large. As a consequence, such buildings cannot store thermal energy over long horizons. For those buildings, an MFPH appears in the TI formulation, while it was absent in the TD envelope. Even though the existence of an \ac{MFPH} may seem conservative, it seems reasonable to restrict the power consumption planning of light and poorly insulated buildings to few hours ahead. 

\subsection{Multi-Dimensional System: A 9-room Building}

In multi-room buildings, thermal exchanges between adjacent rooms exist and may impact the flexibility of the building. Section~\ref{sec:operationalPotentialMultiD} presents two TI flexibility envelope formulations for multi-room buildings: Section~\ref{subsec:energyDecoupled} assumes independent rooms, e.g., representing apartments in a multi-family building, while Section~\ref{subsec:coupledFormulation} considers that the rooms are centrally managed, e.g., describing a single-family house. We assess on the 9-room building presented in Section~\ref{subsec:caseStudyMultiD} the performances of both formulations. 

\begin{figure}
     \begin{subfigure}[b]{\columnwidth}
         \hspace{-0.2cm}
         \includegraphics[width=\columnwidth]{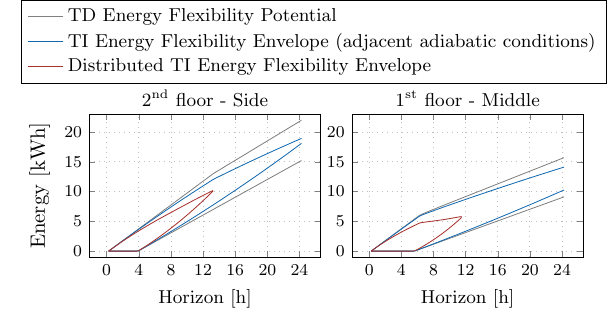}
         \caption{Without supplementary indoor insulation.}
         \label{subfig:withoutInsulation}
         \vspace{0.2cm}
     \end{subfigure}
     \begin{subfigure}[b]{\columnwidth}
         \hspace{-0.25cm}
         \includegraphics[width=0.97\columnwidth]{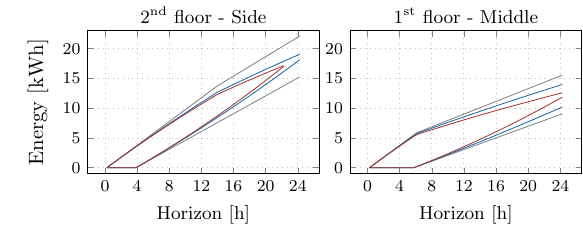}
         \caption{With supplementary indoor insulation.}
         \label{subfig:withInsulation}
     \end{subfigure}
        \caption{Comparison of envelope formulations for two rooms of the 9-room building, with and without supplementary indoor insulation.}
        \label{fig:multiFamilyRoom}
\end{figure}

When the different rooms are considered independent, one TI flexibility envelope is derived per room. Fig.~\ref{fig:multiFamilyRoom} compares, at the room level, three flexibility envelope formulations: the TD energy flexibility envelope, the TI energy flexibility envelope, when it is assumed that the adjacent rooms intrinsically follow the same temperature trajectory as the considered room and therefore no heat exchange would occur with these adjacent rooms, and the TI envelope introduced in Section~\ref{subsec:energyDecoupled}, used in a distributed setting. In the case of no adjacent exchanges, the uni-dimensional TI flexibility formulation is applicable, as rooms are only affected by thermal losses to the outside environment. We specifically investigate two rooms: one on the top-floor with several facade walls, characterized by large thermal losses to the ambient environment, and the other located in the middle of the building and characterized by low thermal exchanges with the ambient environment. 

As Fig.~\ref{fig:multiFamilyRoom} depicts, more energy is necessary to heat up the top-floor room, which is due to large thermal losses to the outside. Under adjacent adiabatic conditions, the TI envelope of this room deviates significantly from its TD envelope, specifically for long horizons. This aligns with previous observations regarding poorly insulated buildings. On the contrary, the middle-floor room is less exposed to thermal losses to the outside, resulting in a lower energy consumption and less differences between the TI and TD envelopes under adjacent adiabatic conditions. 

In practice, adjacent room temperatures differ. Hence, we must employ a TI envelope formulation for multi-dimensional systems. In a distributed setup, we obtain one TI envelope per room. In such case, we must assume the worst-case thermal impact from adjacent rooms, i.e., the power consumption trajectory that causes extreme thermal exchanges to the room. Consequently, if the rooms are strongly coupled, i.e., without supplementary indoor insulation, the multi-dimensional distributed TI envelope significantly deviates from the TI envelope with adjacent adiabatic exchanges. 
However, both envelopes are similar if additional insulation is installed between the rooms, as shown in Fig.~\ref{subfig:withInsulation}.
In this second case, thermal exchanges from adjacent rooms have a limited impact on the energy consumption of individual rooms. This example highlights that a multi-dimensional distributed TI envelope formulation performs well for weakly physically coupled systems.

\begin{figure}
    \hspace{-0.5cm}
    \includegraphics[width=\columnwidth]{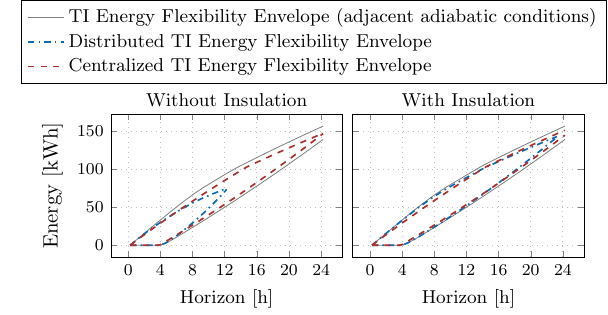}
    \caption{Comparison of TI envelopes for the total flexibility of the 9-room building, with and without supplementary indoor insulation.}
    \label{fig:multiFamilyWithAndWithoutInsulation}
\end{figure}

Poor indoor insulation between rooms may also occur in a single-family building. In such a case, individual rooms are not individual flexibility providers, but provide instead flexibility in an aggregated manner, i.e., they are centrally managed. Therefore, the multi-dimensional centralized TI flexibility envelope formulation is applicable. Fig.~\ref{fig:multiFamilyWithAndWithoutInsulation} represents the total flexibility envelope of the building, i.e., the sum of individual envelopes\footnote{Individual envelopes can be summed up into one total envelope because all heat sources have the same power rating. In case of different power ratings, advanced methods based on Minkowski sums (see e.g. \cite{WEN2024110628}) are required to obtain the total energy envelope.}. It features the total TI envelope, assuming adjacent adiabatic conditions, the total distributed TI envelope, and the total centralized TI envelope. Without additional indoor insulation, the centralized TI envelope significantly differs from the distributed one, but is comparable to the TI envelope, assuming adjacent adiabatic conditions. With additional indoor insulation, all envelopes look alike. As a conclusion, strongly physically coupled systems, such as rooms in a single-family house, should be operated centrally in order to increase their TI flexibility. 

\section{Conclusion}
\label{sec:conclusion}

The TD flexibility potential describes the maximum and minimum energy that a flexible system can consume over a fixed horizon. However, for systems affected by state-dependent losses, power consumption trajectories included in TD envelopes may not fulfill state constraints, leading to unsatisfactory system behavior or infeasibilities. Therefore, this paper introduces TI energy flexibility bounds that guarantee the state constraints' satisfaction for all power trajectories inscribed in the envelope. TI energy envelope formulations are developed for both uni and multi-dimensional systems. 

The proposed TI energy envelope formulations are tested using the example of a heating system in a building, which describes a system with state-dependent losses. This case study reveals some key insights. First, in systems affected by large state-dependent losses, e.g., light constructions or poorly insulated buildings, TI envelopes significantly differ from TD envelopes. In fact, TD envelopes could lead to severe violations of the temperature constraints, and the use of TI envelopes is necessary to guarantee inhabitants' comfort. Additionally, for systems with multiple strongly physically coupled states, e.g., heating systems in a single-family house, it is preferable to manage loads centrally and offer an aggregated TI envelope. However, loads can offer their flexibility independently in multi-dimensional systems with weak physical coupling between the states, e.g., individual heating systems in a multi-family house.  

Future works should analyze more realistic case studies, specifically including the impact of solar irradiation. Besides, our case study focuses on building thermal dynamics. Yet, TI energy envelopes developed in this paper can also find applications to describe the flexibility of other energy-constrained systems affected by self-losses, e.g. lithium-ion batteries with self-losses \cite{Zimmerman2004}. Furthermore, future works may seek additional TI energy envelope formulations, especially optimal ones, i.e., formulations describing the largest TI flexibility energy envelope.

\bibliographystyle{ieeetr}
\bibliography{reference.bib}

\end{document}